\documentclass[aps,pra,twocolumn,showpacs,superscriptaddress]{revtex4-1}
													
\usepackage{amsmath}
\usepackage{amsthm}
\usepackage{latexsym}
\usepackage{amssymb}
\usepackage{bm}
\usepackage{bbm}
\usepackage{appendix}
\usepackage{graphics,epstopdf}
\usepackage{color}
\usepackage[colorlinks=true,linkcolor=blue,citecolor=red,plainpages=false,pdfpagelabels]{hyperref}
\usepackage[normalem]{ulem}

\usepackage{newlfont}
\usepackage{amsfonts}
\usepackage{amsthm}
\usepackage{graphicx}
\usepackage{epsfig}

\usepackage[thinc]{esdiff}

\usepackage{times}

\newcommand{\ket}[1]{|#1\rangle}

\DeclareOldFontCommand{\rm}{\normalfont\rmfamily}{\mathrm}

\begin{document}
\title{Stronger Quantum Speed Limit }
\author{Dimpi Thakuria}\email{dimpithakuria@hri.res.in}

\affiliation{\(\)Harish-Chandra Research Institute,  A CI of Homi Bhabha National Institute, Chhatnag Road, Jhunsi, Prayagraj  211019, Uttar Pradesh, India
}
\author{Arun Kumar Pati}\email{akpati@hri.res.in}
\affiliation{\(\)Harish-Chandra Research Institute,  A CI of Homi Bhabha National Institute, Chhatnag Road, Jhunsi, Prayagraj  211019, Uttar Pradesh, India
}

\begin{abstract}
The quantum speed limit provides fundamental bound on how fast a quantum system can evolve between the initial and the final states. 
For the unitary evolution, the celebrated Mandelstam-Tamm (MT) bound has been widely studied for various systems. Here, we prove a stronger quantum speed limit (SQSL) for all quantum systems undergoing arbitrary unitary evolution and show that the MT bound is a special case of the stronger quantum speed limit. We apply our result for single system as well as for composite systems in separable and entangled states and show that the new bound is indeed tight. The stronger quantum speed limit will have wide range of applications in quantum control, quantum computing and quantum information processing.
	\end{abstract}
	\maketitle
	
	\section{Introduction}
	\label{sec1}
     Uncertainty principles and uncertainty relations have played a vital role in the formulation of quantum mechanics which helped us to unveil the behavior of the microscopic world in many different ways. The very first uncertainty principle discovered by Heisenberg suggested a lower bound to the product of the error and disturbance for two canonically conjugate quantum mechanical variables~\cite{Heisenberg1927}. On the other hand, the uncertainty relations are capable of capturing the inherent restrictions in preparation of quantum systems for which precise measurement of two conjugate quantities can be carried out~\cite{Wheeler1983}. This interpretation was quite fruitful for the uncertainty relations like position-momentum, angular position-angular momentum uncertainty relations etc. However, the energy-time uncertainty relation has generated debates since the early days of quantum mechanics as time is not an operator \cite{JHi1996,JHi1998,JHi2002}.
    
 Robertson established an uncertainty relation  for any two quantum-mechanical observables, not necessarily the conjugate variables~\cite{Robertson1929}. This relation quantitatively expresses the impossibility of jointly sharp preparation and therefore measurement, of any two incompatible observables.  However, the Robertson uncertainty relation do not completely express the
 incompatibility nature of two non-commuting observables. The stronger uncertainty relations have been proved which capture the notion of 
 incompatibility more stringently \cite{Maccone2014}.
 This brought more clarity to the role played by the uncertainty relations in quantum mechanics. This also provided a firm ground to the already existing uncertainty relations like the position-momentum uncertainty relation. But time, not being an observable, energy-time uncertainty relation still lacked a valid proof and an unambiguous interpretation.

  Mandelstam and Tamm (MT) derived an energy-time uncertainty like relation which essentially follows from the Robertson uncertainty relation \cite{Mandelstam1945}. For a quantum system evolving under the Hamiltonian $H$, the MT relation
    is given by
    \begin{eqnarray}
    \label{1}
    \Delta H T\geq\hbar cos^{-1}(|\langle\Psi(0)|\Psi(T)\rangle|),
    \end{eqnarray}
    where \(|\Psi(0)\rangle\) is the initial state at $t=0$, \(|\Psi(T)\rangle\) is the final state after $t=T$ and \(\Delta H\) is the instantaneous  uncertainty in the energy. Clearly, relation (\ref{1}) possess a similar form as of the energy-time uncertainty relation, but the uncertainty in time is strictly replaced by the evolution time, $T$. Thus the bound (\ref{1}) captures the intrinsic time scale of the system rather than the uncertainty in measurement of time \cite{S.Deffner2017,Busch2008,Busch1990,P.Busch1990-1}. The above relation yields the lower bound on the evolution time as given by
    \begin{eqnarray}
    \label{2}
    T\geq \frac{\hbar cos^{-1}(|\langle\Psi(0)|\Psi(T)\rangle|)}{\Delta H}.
    \end{eqnarray}
 The relation(\ref{2}) suggests a minimal time within which a system can go from an initial state to a final state. In other words, it provides a limit to how fast a quantum system can evolve. Thus, the lower limit is often known as the quantum speed limit (QSL). Later, Anandan and Aharonov  derived the MT bound using the geometric approach of quantum evolution and the notion of  the speed of quantum system on the projective Hilbert space ${\cal P}(\cal H)$ \cite{Anandan1990}. The speed of quantum evolution was also defined using the Riemannian metric in Ref.\cite{Pati1991}. \par
    
    The relation (\ref{2}) captures the minimum time interval over which a quantum system undergoes unitary evolution from an initial state to a final state. For the cases, where the inequality holds tight, the QSL is nothing but the evolution time itself. In many problems of physics, the Fermi's golden rule and the perturbation theory are often applied for the calculation of the transition rate which can give transition time. These are certainly important, valid under certain approximations and become computationally inconvenient under some conditions ~\cite{Boykin_2007}. 
On the other hand, the relation (\ref{2}) is appropriate under any condition for a unitary evolution. It expresses the evolution time in terms of energy width $\Delta H$ and the overlap between the initial and the final state of the system, which are comparatively easy to calculate. 
    
In recent years, the notion of quantum speed limits have been very useful in providing us with the maximal rate of quantum information processing \cite{Lloyd1981}, the maximal rate of quantum communication \cite{Bekenstein1981}, the maximal rate of quantum entropy production \cite{Deffner2010,Das2018} and various other directions ~\cite{Mohan2021,Campaioli2022,Pandey2022}. QSL has a wide range of applications in quantum metrology~\cite{Campbell2018}, quantum thermodynamics~\cite{Mukhopadhyay2018}, and optimal control theory~\cite{Caneva2009,Campbell2017} and so on. Undoubtedly, the quantum speed limit has generated lot of interest in the physics community over last several decades. \par
   In the literature, there are several other approaches to obtain quantum speed limits. For example, the Margolus-Levitin bound involves the average of energy in the quantum speed limit \cite{Margolus1998,Levitin2009}. The quantum speed limit has also been generalized for mixed states undergoing unitary evolution \cite{Uhlmann1992,Uffink1993,Margolus1998,PATI1999,Giovannetti_2003,Giovannetti_2004,Luo2005,Batle2005,Andrews2007,Zander2007,Zielinski2006,Kupferman2008,Levitin2009,Yurtsever2010,Chau2010,Frowis2012,Ashhab2012,Taddei2013,Mondal2016} and open system dynamics \cite{S.Deffner2013, Campo2013, Taddei2013, Fung2013, Pires2016, S.Deffner2020,Jing2016,Pintos2021,Mohan2022,Mohan21,Dimpi2022}.
    In this paper, we seek a stronger form of quantum speed limit. We show that the new bound provides a tighter expression compared to the existing bounds such as the MT bound. The lower bound for the evolution time is more accurate in predicting the evolution of a quantum system than the existing bounds. We also show that the new bound is more versatile than the existing bounds in the sense that by suitable choice of an orthogonal state to the system state, we can make the lower bound for the evolution time more tighter. This option is not available for the MT bound. We apply our result for composite systems which are prepared in the separable state and 
    entangled states and obtain the quantum speed limits in the presence and absence of interaction Hamiltonian. We find that the stronger quantum speed limit (SQSL) is indeed tight compared to MT bound in almost all cases.
    
    The present paper is organised as follows. In Section II, we derive the stronger quantum speed limit. In Section III, we analyse the obtained bound for the non-interacting systems which are prepared in separable and entangled states, respectively. In Section IV, we consider an initial entangled state that evolves under interaction Hamiltonian and find that the new bound surpasses the MT bound. Finally, in Section V we summarize our results.

    \section{Stronger quantum speed limit}
    \label{sec2}
    Consider a quantum system with state vector $|\Psi(t)\rangle \in \cal{H}^N$. Let it evolve under the Schr\"odinger equation $$i\hbar\dfrac{d}{dt}|\Psi(t)\rangle = H|\Psi(t)\rangle $$
    over a time interval $t: [ 0,  T]$ with $\ket{\Psi(0)}$ being the initial state and $\ket{\Psi(T)}$ being the final state.
    Now, we would like to prove the stronger quantum speed limit which can capture the evolution time more precisely compared to the MT bound. To prove that stronger quantum speed limit, we consider one of the stronger uncertainty relations proved by  Maccone and Pati \cite{Maccone2014}. For any two observables \(A\) and \(B\), this uncertainty relation can be expressed as
    
    \begin{eqnarray}
    \label{3}
    \Delta A\Delta B \geq \pm\frac{i}{2} \langle\Psi(t)|[A,B]|\Psi(t)\rangle + R(t)\Delta A\Delta B 
    \end{eqnarray}
    \begin{eqnarray}
    where\quad
    R(t) =\frac{1}{2}\left|\langle\Psi^\perp(t)|\frac{A}{\Delta A}\mp i\frac{H}{ \Delta H}|\Psi(t)\rangle\right|^2\nonumber,
    \end{eqnarray}
    \(\Delta A=\sqrt{\langle A^2 \rangle-\langle A \rangle^2}\) and \(\Delta B=\sqrt{\langle B^2 \rangle-\langle B \rangle^2}\) are the standard deviation measures of uncertainty for the observables \(A\) and \(B\), respectively. Note that \(|\Psi^\perp (t)\rangle\) is an arbitrary state orthogonal to the state, \(|\Psi (t)\rangle\) of the system. The pair of signs \((\pm,\mp)\) is chosen in such a way that the term   \(\pm\frac{i}{2}\langle[A,B]\rangle\) remains positive. This relation can also be seen as a modified version of the Robertson uncertainty relation with the extra term, $R(t)$ present in it. Since $R(t)$ is a positive quantity, this relation is tighter than the Robertson uncertainty relation. \par
    
    Since the MT bound is a consequence of the Robertson uncertainty relation, it is natural to ask whether one can derive a stronger quantum speed limit bound from the above relation (\ref{3}). If one of the operators, say $B$ is chosen to be the Hamiltonian of a system, then  using the equation of motion for the average of the observable  
    \begin{eqnarray}
    \label{4}
   i \hbar\dfrac{d\langle A \rangle}{dt}=\left\langle[A,H]\right\rangle
    \end{eqnarray}
    we have 
    \begin{eqnarray}
    \label{5}
    \Delta A\Delta H(1-R(t))\geq \frac{\hbar}{2}\left|\frac{d \langle A \rangle}{d t}\right|.
    \end{eqnarray}
 Consider $A=|\Psi (0)\rangle\langle\Psi (0)|$ as the projection operator defined for the initial state of the system. Since $A^2=A$, the variance becomes $\Delta A^2=\langle A \rangle-\langle A \rangle^2$, where $\langle A \rangle=|\langle\Psi(0)|\Psi(t)\rangle|^2$ is nothing but the transition probability between the initial and the state at time, t. Since \(|\langle\Psi(0)|\Psi(t)\rangle|=cos\frac{s_{0}(t)}{2}\), we can write \begin{eqnarray}
    \label{6}
    \Delta A=\frac{1}{2} sin(s_{0}(t)), \dfrac{d \langle A \rangle}{d t}=-\frac{1}{2} sin (s_{0}(t)) \dfrac{d s_{0}(t)}{d t}.
    \end{eqnarray}
    Using these expressions, relation (\ref{5}) can be re-written as
    \begin{eqnarray}
    \label{7}
    \begin{aligned}
    \Delta H(1-R(t)) & \geq\frac{\hbar}{2}\dfrac{d s_{0}(t)}{d t}.
     \end{aligned}
    \end{eqnarray}
    If we consider the Hamiltonian $H$ to be time independent, then \(\Delta H\) is also a time independent quantity. Performing integration in both side of the inequality (\ref{7}) with respect to time yields
     \begin{eqnarray}
    \label{8}
    T  \geq \frac{\hbar S_0}{2\Delta H}+\int_{0}^{T}R(t)dt,
    \end{eqnarray}
 \begin{eqnarray}
     where\quad
     R(t)=\frac{1}{2}\left|\langle\Psi^\perp(t)|\frac{A}{\Delta A}-i\frac{H}{\Delta H}|\Psi(t)\rangle \right|^2\nonumber
    \end{eqnarray}
    and $T$ is the time over which the system evolves from the initial state to the final state. Here,  \(S_0=2cos^{-1}(|\langle\Psi(0)|\Psi(T)\rangle|)\) is the geodesic distance between the initial and the final state on the projective Hilbert space, ${\cal P}({\cal H})$ as measured by the Fubini Study metric \cite{Cheng2010,Braunstein1994}. Expression (\ref{8}) gives the new expression for quantum speed limit
    \begin{eqnarray}
    \label{10}
    T_{SQSL}=\frac{\hbar S_0}{2\Delta H}+\int_{0}^{T}R(t)dt.
    \end{eqnarray}
    The bound involves an additional term \(\int_{0}^{T}R(t)dt\) along with the MT bound (\ref{2}). Since $R(t) \geq 0$, 
   the bound (\ref{8}) is stronger than the MT bound. This is the central result of our paper and we call $T_{SQSL}$ as the stronger 
   quantum speed limit bound. Note that the SQSL bound given in Eq (\ref{8}) can be compactly expressed as 
   $$T\geq \frac{\hbar \Gamma S_{0}}{2\Delta H} $$ 
   where $\Gamma=\frac{1}{1-\overline{R(t)}}$ and $\overline{R(t)}=\frac{1}{T}\int_{0}^{T}R(t)dt$ is the time-average of the quantity $R(t)$. Now, one can ask under what condition $R(t)$ vanishes? Below, we answer this question (For simplicity, we assume $\hbar=1$).\\
\\$Proposition \quad 1$. When a quantum system evolves along the shortest geodesic path, the term $R(t)$ vanishes.\\
\\$Proof$: Consider the geodesic equation for pure state~\cite{Pati1994} evolving under a time independent Hamiltonian, $H$ 
   \begin{equation}
   \label{10-1}
   |\overline{\Psi}(t)\rangle=\cos{\Delta H t} |\overline{\Psi}(0)\rangle+\frac{sin{\Delta H t}}{\Delta H}|\Dot{\overline{\Psi}}(0)\rangle.
   \end{equation}           
   Here, $|\overline{\Psi}(t)\rangle$ is the parallel-transported vector defined by  $|\overline{\Psi}(t)\rangle=e^{i\int_{0}^{t}\langle H\rangle dt}|\Psi(t)\rangle$. The velocity vector then can be written as
   \begin{eqnarray}
   \label{10-2}
   |\Dot{\overline{\Psi}}(t)\rangle=-\Delta H(\csc{\Delta H t} |\overline{\Psi}(0)\rangle-\cot{\Delta H t}|\overline{\Psi}(t)\rangle). 
  \end{eqnarray}
  Using the definition of $|\overline{\Psi}(t)\rangle$, we express $R(t)$ as 
  \begin{equation}
  \label{10-3}
  R(t)=\frac{1}{2} \left|\langle\overline{\Psi}^\perp(t)|\left(\frac{A}{\Delta A}|\overline{\Psi}(t)\rangle +\frac{1}{ \Delta H}|\Dot{\overline{\Psi}}(t)\rangle\right) \right|^2,
  \end{equation}
  where $A=|\overline{\Psi}(0)\rangle\langle\overline{\Psi}(0)|$ and $\langle A\rangle=\cos^2{\frac{s_0(t)}{2}}$. Now, to find $R(t)$ along the geodesic, we put Eq.(\ref{10-1}), (\ref{10-2}) in Eq.(\ref{10-3}). Further applying the parallel-transport condition $\langle\overline{\Psi}(t)|\Dot{\overline{\Psi}}(t)\rangle=0$, we get
  \begin{equation}\label{10-4}
  R(t)=\frac{1}{2} \left|\left(\frac{\cos{\Delta H t}}{\Delta A}-\frac{1}{\sin{ \Delta H t}}\right) \langle\overline{\Psi}^\perp(t)|\overline{\Psi}(0)\rangle\right|^2. 
  \end{equation}
 Since $\Delta A=\frac{1}{2}\sin{2\Delta H t}$, Eq. (\ref{10-4}) yields $R(t)=0$. Hence the proof.

 This shows that when the quantum system evolves along the geodesic, the stronger quantum speed limit coincides with the MT bound.
 Next, one may ask what happens to the evolution path when $R(t)$ is zero. We answer this below.
   \\
   $Proposition\quad 2$. If $R(t)=0$ and $\dfrac{ds}{dt}=\dfrac{ds_0}{dt}$, the system evolves along the geodesic path.\\
   \\$Proof$: Consider the first condition $R(t)=0$. Using Eq.(\ref{10-3}), it can be written as
   \begin{eqnarray}
   \label{10-4a}
  \langle\overline{\Psi}^\perp(t)|\left(\frac{A}{\Delta A}|\overline{\Psi}(t)\rangle +\frac{1}{ \Delta H}|\Dot{\overline{\Psi}}(t)\rangle\right)=0.
  \end{eqnarray}
 Since  $|\overline{\Psi}^\perp(t)\rangle$ is arbitrary, the condition (\ref{10-4a}) must hold for all possible $|\overline{\Psi}^\perp(t)\rangle$. This is possible only when
 \begin{eqnarray}
 \label{10-5}
 \begin{aligned}
 \frac{A}{\Delta A}|\overline{\Psi}(t)\rangle +\frac{1}{ \Delta H}|\Dot{\overline{\Psi}}(t)\rangle &\propto |\overline{\Psi}(t)\rangle\\
 &=D |\overline{\Psi}(t)\rangle,
 \end{aligned}
 \end{eqnarray}
where $D$ is a proportionality factor. Now, taking the inner product with respect to $|\overline{\Psi}(t)\rangle$  in both sides of Eq. (\ref{10-5}) and using the parallel transport condition, $\langle\overline{\Psi}(t)|\Dot{\overline{\Psi}}(t)\rangle=0$ we obtain, $D=\frac{\langle A \rangle}{\Delta A}$. Using this expression in Eq.  (\ref{10-5}) we have
   \begin{eqnarray}
   |\Dot{\overline{\Psi}}(t)\rangle =\frac{\Delta H}{\Delta A}\left(\langle A \rangle-A\right)|\overline{\Psi}(t)\rangle.
   \end{eqnarray}
  Now, differentiating both sides of the above equation with respect to time yields  
  \begin{eqnarray}
  \label{10-f}
  |\ddot{\overline{\Psi}}(t)\rangle =\Delta {H}^2P
  _1|\Psi(t)\rangle
  +P_2\langle\overline{\Psi}(0)|\overline{\Psi}(t)\rangle|\overline{\Psi}(0)\rangle,
  \end{eqnarray}
  \begin{eqnarray}
  \label{10g}
  \begin{aligned}
where\quad P_1&=\cot^2{\frac{s_0(t)}{2}}-\frac{\csc^2{\frac{s_0 t}{2}}}{2\Delta H}\dfrac{ds_0}{dt}\\and\quad
  P_2&=\frac{\cos{s_0 t}}{\sin^2{s_0 t}}\left(\dfrac{ds_0}{dt}-2\Delta H\right).
   \end{aligned}
  \end{eqnarray}
 Then we apply the second condition of the proposition,i.e., $ds_0/dt=ds/dt$. Here $ds/dt$ is the evolution speed of the system, which is well defined for unitary evolution~\cite{Anandan1990} as $ds/dt=2\Delta H$. Applying this condition in Eq.(\ref{10g}), we get $P_1=1$ and $P_2=0$. Therefore, Eq.(\ref{10-f}) gives
 \begin{eqnarray}
 \label{10h}
  |\ddot{\overline{\Psi}}(t)\rangle=-\Delta {H}^2|\Psi(t)\rangle.
  \end{eqnarray}
This is nothing but the geodesic equation \cite{Pati1994} followed by the system evolving under time independent Hamiltonian $H$. For derivation of the geodesic equation using variational methods reader may refer~\cite{MUKUNDA1993,Pati1994}.\par 
In the above, the bound is discussed in the general context. Although it is tighter than the MT bound, its performance may vary with the system under consideration. In the following sections, we discuss how to obtain a tighter quantum speed limit for various types of systems, including interacting and non-interacting subsystems.\par
    
    \section{The stronger quantum speed limit for non-interacting subsystems}
     We now illustrate the new bound for the interaction free subsystems. We carry out the discussion in two parts. First, we discuss the SQSL for initial product state and then we discuss for entangled state.
    \subsection{SQSL for Separable State}
    Consider a system with $M$ number of subsystems  initially at $t=0$, in the product state \cite{Giovannetti_2004} 
    \begin{eqnarray}
    \label{12}
    |\Psi(0)\rangle=|\psi(0)\rangle_1 |\psi(0)\rangle_2....|\psi(0)\rangle_M, 
     \end{eqnarray}
     where $|\psi(0)\rangle_i=\frac{1}{\sqrt{N}}\sum_{n=o}^{N-1} |n\rangle_i (i=1,2....M)$. Here, $N$ is the dimension of the Hilbert space of each subsystem. The system  now evolves under the Hamiltonian \(H=\sum_{i=1}^{M}H_i\), where
    \begin{eqnarray}
    \label{14}
    H_i=\omega\hbar\sum_{n=0}^{N-1}n|n\rangle\langle n|.
    \end{eqnarray}
    This is a homogeneous energy system where every subsystem has an energy expectation value
    \begin{eqnarray}
    \label{15}
    \langle H\rangle=\frac{ \hbar \omega M (N-1)}{2}
    \end{eqnarray}
    and the uncertainty in energy
    \begin{eqnarray}
    \label{16}
    \Delta H =\frac{\hbar\omega \sqrt{M(N^2-1)}}{2\sqrt{3}}.
    \end{eqnarray}
    After time $t$, the transition amplitude of the system becomes
    \begin{eqnarray}
    \label{17}
    \langle \Psi(0)|\Psi(t)\rangle=\left(\frac{1-e^{-i\omega t N}}{N(1-e^{-i\omega t})}\right)^M.
    \end{eqnarray}
    Thus, the transition probability is given by
    \begin{eqnarray}
    \label{18}
    |\langle \Psi(0)|\Psi(t)\rangle|^2=\left|\frac{1}{N}\frac{sin(\frac{\omega t N}{2})}{sin(\frac{\omega t}{2})}\right|^{2M}.
    \end{eqnarray}
    Using the relations (\ref{18}), (\ref{16}) and (\ref{2}) we find the MT quantum speed limit for the above system 
    \begin{eqnarray}
    \label{20}
    T_{QSL}^{MT}=\frac{S_0}{2\Delta H}=2\sqrt{3}\frac{cos^{-1}\left|\frac{1}{N}\frac{sin(\frac{\omega T N}{2})}{sin(\frac{\omega T}{2})}\right|^M}{\sqrt{M(N^2-1)}}.
    \end{eqnarray} 
    Here, we have assumed $\hbar=1$ for the simplicity of the calculation. To obtain the new quantum speed limit time (\ref{10}), the additional term \(\int_{0}^{T}R(t)dt\) has to be calculated. To specify \(R(t)\), a choice of orthgonal state \(|\Psi^\perp (t) \rangle\) is required. In order to yield a promising expression of quantum speed limit time, it is necessary to choose \(|\Psi^\perp (t) \rangle\) properly. In the following, we analyse the bound (\ref{10}) for a particular choice of the orthgonal state:
    \begin{eqnarray}
    \label{21}
    |\Psi^\perp(t)\rangle =\frac{A-\langle A \rangle }{\Delta A}|\Psi(t)\rangle,
    \end{eqnarray}
    where $A=|\Psi(0)\rangle\langle\Psi(0)|$. Inserting the above orthogonal state in (\ref{8}) the expression of \(R(t)\)  can be re-written as 
    \begin{eqnarray}
    \label{22}
    R(t)
    =\frac{1}{2}\left|1-i\frac{\langle AH\rangle-\langle A\rangle\langle H\rangle}{\Delta H\Delta A}\right|^2,
     \end{eqnarray}
   where \(\langle A \rangle\), \(\Delta A\) can be easily calculated using (\ref{6}) and (\ref{18}). Simple calculation shows that  
    \begin{eqnarray}
    \label{23}    
     \langle AH \rangle=\frac{M\langle \Psi(0)| \Psi(t) \rangle \langle \psi(0)|_iH_i|\psi(t)\rangle_i }{\langle \Psi(0)|\Psi(t) \rangle^{1/M}}, 
    \end{eqnarray}
    where
    \begin{eqnarray}
    \label{24}
    \begin{aligned}
    \langle \psi(0)|_i H_i|\psi(t)\rangle_i &=\frac{\sum_{n=0}^{N-1} n e^{-in\omega t}}{N}\\
     &=\frac{e^{-i\omega t}}{Z_1}\left[\frac{1-Z_2}{Z_1}-(N-1)Z_2\right] \nonumber,\\
    Z_1=1-e^{-i \omega t}, Z_2 &=e^{-i(N-1)\omega t}. 
     \end{aligned}
    \end{eqnarray}
    \begin{figure}[htp]
    \centering   
    \includegraphics[width=8cm]{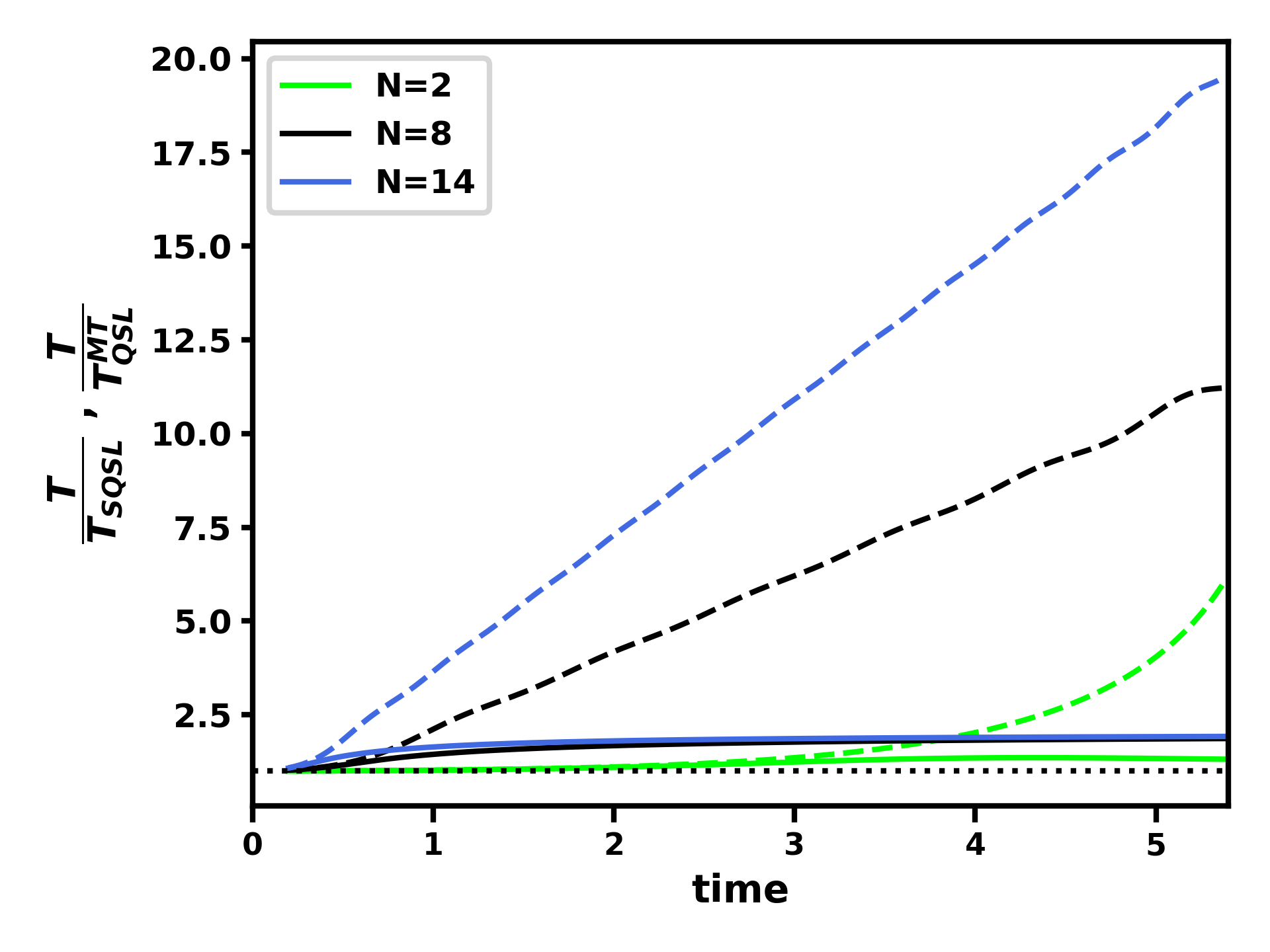}
    \caption{\(T/T_{SQSL}\) and \(T/T^{MT}_{QSL}\) are plotted against time for different values of $N$ by fixing the value of $M$ as \(M=2\). The MT bounds are represented by the dashed lines. The new bounds are represented by the continuous lines. The dotted line represents $T/T_{SQSL}=T/T^{MT}_{QSL}=1$.}
    \label{fig:1}
    \end{figure}
    \par
    \begin{figure}[htp]
    \centering  
    \includegraphics[width=8cm]{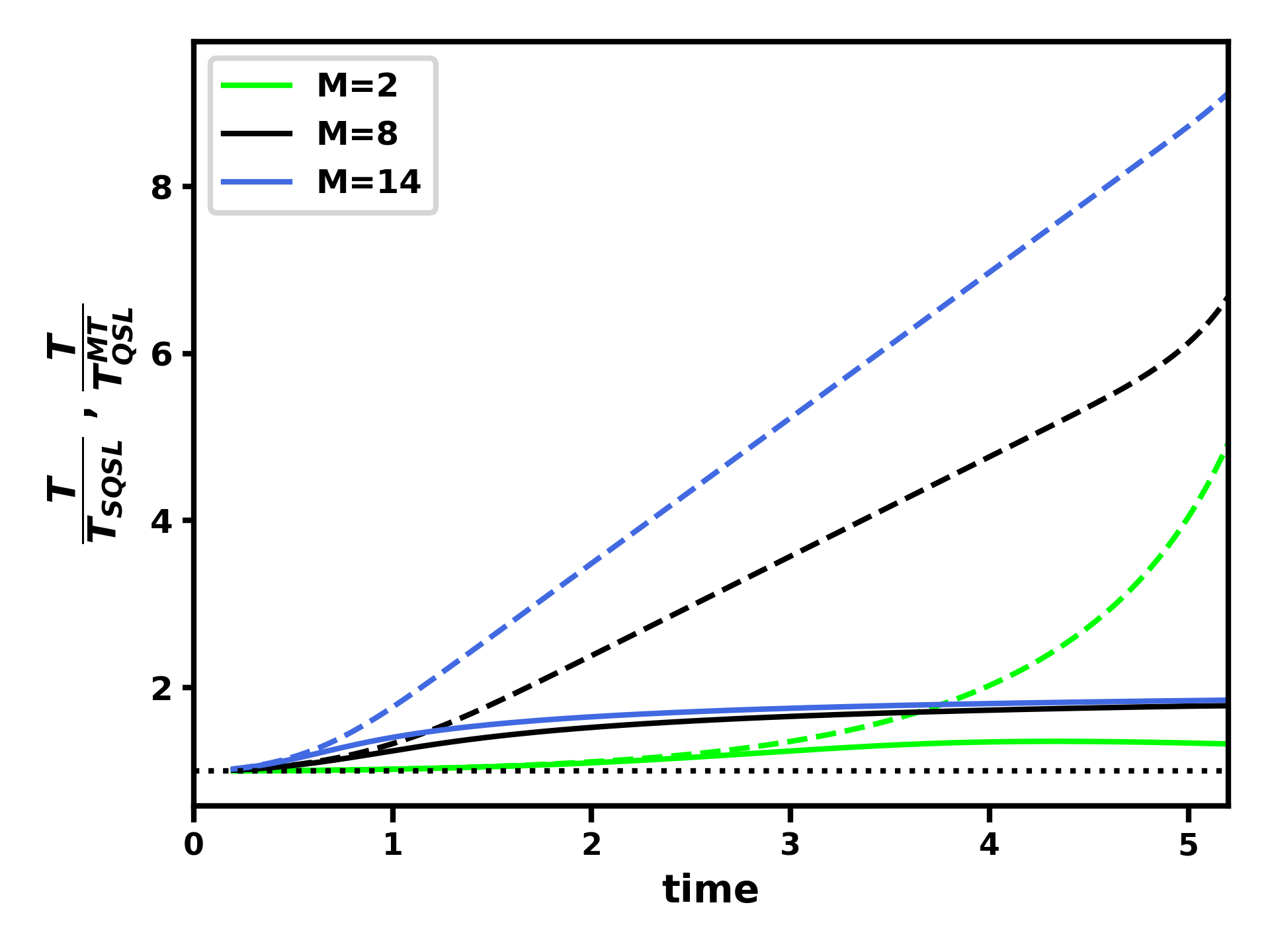}
    \caption{\(T/T_{SQSL}\) and \(T/T^{MT}_{QSL}\) are plotted against time for different values of $M$ by fixing the value, \(N=2\). MT bounds are represented by the dashed lines and that for the new bounds are represented by the continuous lines. The dotted line represents $T/T_{SQSL}=T/T^{MT}_{QSL}=1$.}
    \label{fig:2}
    \end{figure}
    Assembling all the above expressions we can find $R(t)$ and putting it in the Eq (\ref{10}), we obtain the new bound on the evolution time.\par
   To analyse the new bound, the ratio $T/T_{SQSL}$ is observed with respect to the evolution time. We focus on two aspects of the new bound: first, how accurately the bound tells about the evolution time of the system and second, how well the bound performs over the existing bound like MT bound. The former can be inferred by observing the closeness between $T/T_{SQSL}$ and 1. The more $T/T_{SQSL}$ is close to 1, the more accurate the bound is about the evolution time. When $ T/T_{SQSL}=1 $, the bound ($\ref{10}$) gives the exact value of the evolution time i.e, $T=T_{SQSL}$. For the later purpose we have to check the tightness of the new bound relative to the MT bound. In the previous section, the new bound is proved to be tighter than the MT bound, i.e., $T_{SQSL}\geq T_{QSL}^{MT}$. For any evolution time $T$, this condition can be rewritten as  
    \begin{eqnarray}
    \label{26}
    \frac{T}{T_{QSL}^{MT}} \geq \frac{T}{T_{SQSL}} 
    \end{eqnarray}
    which also implies
    \begin{eqnarray}
    \label{27}
    \left |\frac{T}{T_{QSL}^{MT}}-\frac{T}{T_{SQSL}} \right | & \geq 0.
    \end{eqnarray}
    Therefore, the difference $\delta=\left|\frac{T}{T_{QSL}^{MT}}- \frac{T}{T_{SQSL}}\right|$ can be considered as a quantitative measure of the relative tightness between the two bounds. A greater value of $\delta$ implies a greater relative tightness between them.\par
   A numerical estimation of $T/T_{SQSL}$ is made for the range of evolution time, $0$ to $\frac{2\pi}{\omega}$. Since the expression of $T_{SQSL}$ for the above system explicitly depends on the number of subsystems, $M$ as well as the dimension of each subsystem, $N$ we can observe the bound for different values of $M$ and $N$.\par
   In the Fig.1, the bound is observed for different values of $N$ of a bipartite system. For any specific value of $N$, the figure shows that most of the time, the ratio \(T/T_{SQSL}\) for the new bound (\ref{10}) remains close to 1. In contrast to that for MT bound, with the increasing evolution time, \(T/T^{MT}_{QSL}\) significantly diverges from 1. Therefore, at any time, the new bound informs more accurately about the evolution time compared to the MT bound. Again the relative tightness, $\delta$ is represented by the gap between the dotted line and the continuous line for each value of $N$. The gap remains nonzero and increases significantly with increasing evolution time. It indicates that the bound (\ref{10}) is significantly tighter than the MT bound at greater evolution times. Similar behaviour is also observed in Fig.2 for any value of subsystems, $M$ when $N$ is fixed. The above observations remain valid for any value of $M$ and any dimension of Hilbert space, $N$ of the individual subsystems. Fig.1 further shows that MT bound loosens up as $N$ increases. Fig.2 shows that the MT bound loosens up for increasing $M$ as well. But in both the cases, the new bound remains significantly tight. All these observations suggest that (\ref{21}) is a good choice of orthogonal state, that yields a stronger bound.\par
   In the above, we have observed the promising nature of the new bound with the particular choice of orthogonal state (\ref{21}). However, this choice of orthogonal state involves only the system's parameters. Once the system is considered, the orthogonal state is automatically chosen by the above form. In the following, we suggest a form of orthogonal state which along with the system's parameters, also depends on some other parameters that are independent of the system. The presence of such parameters provides the freedom to further adjust the orthogonal state and yield an optimal bound on the evolution time for the considered system. For the sake of illustration, we consider a composite system, whose subsystems belong to two dimensional Hilbert spaces.\par
   Let the initial state of the composite system is given by $|\Psi(0)\rangle=|\psi(0)\rangle^{\otimes M}$, where $|\psi(0)\rangle=\alpha|0\rangle+\beta|1\rangle$. The system evolves under the same Hamiltonian as considered in the equation (\ref{14}) with an average energy
  $\langle H \rangle = M |\beta|^2 \omega \hbar$
    and fluctuation in energy $\Delta H = \sqrt{M} \omega \hbar$.
    We choose $\hbar=1$. Now the MT bound for the above system becomes
    \begin{eqnarray}
    \label{32}
    T^{MT}_{QSL}=\frac{\cos^{-1}{|\langle\Psi(0)|\Psi(T)\rangle|}}{\sqrt{M}  |\alpha||\beta|}
    \end{eqnarray}
    where
    \begin{equation}
    \label{33}
    |\langle\Psi(0)|\Psi(t)\rangle|=\left(|\alpha|^4+|\beta|^4+2 |\alpha|^2|\beta|^2 \cos{\omega t}\right)^{M/2}.
    \end{equation}
    In order to obtain a parametric form of the orthogonal state as mentioned earlier, we choose the orthogonal state as
    \begin{eqnarray}
    \label{34}
    |\Psi^\perp(t)\rangle= \frac{O-\langle O\rangle }{\Delta O}|\Psi(t)\rangle, 
    \end{eqnarray}
    \begin{eqnarray}
      O & = O_1\otimes O_2\otimes... O_i\otimes... O_M, \nonumber 
      \end{eqnarray}
     \begin{align*}
     where \quad
    O_i &=\overrightarrow{\sigma_i}.\hat{n_i}\\
    &=\sin{\theta_i} \cos{\phi_i} \sigma_x+\sin{\theta_i} \sin{\phi_i} \sigma_y+\cos{\theta_i}\sigma_z.\nonumber
    \end{align*}
     Here, $\overrightarrow{\sigma_i}$ is the Pauli vector and $\hat{n_i}$ is the unit Bloch vector, corresponding to each qubit. Since $\theta_i$ and $\phi_i$ are the polar and azimuthal angles corresponding to $\hat{n_i}$, $0<\theta_i<\pi$ and $0<\phi_i<2\pi$. There are total $2M$ number of parameters ($\theta_i$,$\phi_i$)  present in the above expression. Each set of $\theta_i$ and $\phi_i$ provides a possible choice of orthogonal state. Thus the form (\ref{34}) as a whole, represents an infinite set of orthogonal states. The remaining task is to find the suitable sets of these parameters that yields an optimal bound on the evolution time.  Optimizing the bound implies minimizing $T/T_{SQSL}$ over the set of $\theta_i$ and $\phi_i$. This requires an expression of $T_{SQSL}$ in terms of $\theta_i$ and $\phi_i$. To make the task simpler, we reduce the number of parameters present in the Eq (\ref{34}) by setting the conditions
    \begin{eqnarray}
    \label{34a}
    \theta_i=\theta \quad ,\quad \phi_i=\phi.
    \end{eqnarray}
    Inserting the expressions  (\ref{34}), (\ref{34a}) in (\ref{8}) we write
    \begin{equation}
    \label{35}
    R(t)=\frac{1}{2}\left|\frac{\langle OA \rangle-\langle O \rangle \langle A \rangle}{\Delta O \Delta A}-i \frac{\langle OH \rangle-\langle O \rangle \langle H \rangle}{\Delta O\Delta H }\right|^2.
    \end{equation}
    Here, the individual terms are given by
    \begin{eqnarray}
    \label{36}
    \begin{aligned}
    \langle O \rangle &= \langle O_i \rangle ^M,\\
    \langle OA \rangle &=\langle O_i A_i \rangle^M,\\ 
    \langle OH \rangle &= M \langle O_i H_i \rangle \langle O_i \rangle^{M-1}, 
     \end{aligned}
     \end{eqnarray}
     where
     \begin{eqnarray}
     \begin{aligned}
      \langle O_i \rangle \quad&=2(A_1\cos{\phi}+A_2\sin{\phi})\sin{\theta}+(|\alpha|^2-|\beta|^2)cos{\theta}, \\
  \langle O_i A_i \rangle&=(|\alpha|^2+|\beta|^2e^{i\omega t}) B_1\sin{\theta}+B_2\sin{\omega t} ,\\
    \langle O_i H_i \rangle&= \alpha^*\beta \sin{\theta} e^{-i(\phi + \omega t)}-|\beta|^2 \cos{\theta},
     \end{aligned}
    \end{eqnarray}
     and
    \begin{eqnarray}
    \begin{aligned}
    A_1&=2Re(\alpha \beta^* e^{i\omega t}),\\
    A_2&=2 i Im(\alpha \beta^* e^{i\omega t}),\\
    B_1&=\alpha^* \beta e^{-i\phi}+\alpha\beta^*e^{i(\phi-\omega t)},\\
    B_2&=|\alpha|^4-|\beta|^4+2i|\alpha|^2|\beta|^2.\nonumber
     \end{aligned}
    \end{eqnarray}
     The notations $Re()$ and $Im()$ represent the real and imaginary part of the respective arguments. Now $\langle A \rangle$, $\Delta A$ can be directly obtained using (\ref{33}).\begin{figure*}
     \centering    
    \includegraphics[width=16cm]{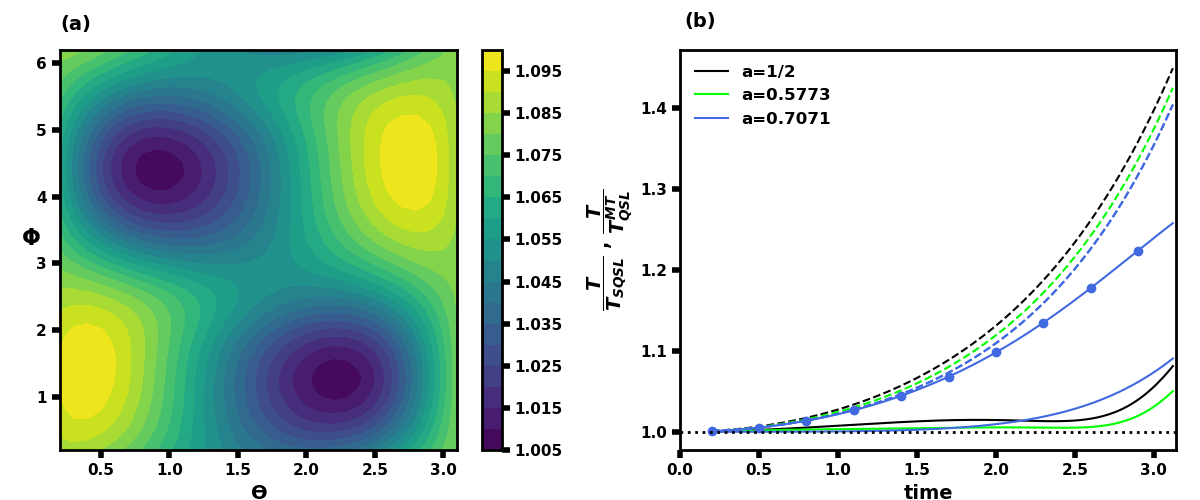}
     \caption{In FIG.(a), $\overline{\frac{T}{T_{SQSL}}}$ is plotted against different values of $\theta$ and $\phi$, for $\alpha=\frac{1}{\sqrt{3}}$ and $M=2$. The optimum pairs of $\theta$ and $\phi$ corresponding to the lowest value of $\overline{\frac{T}{T_{SQSL}}}$ are found to be (4.4,0.9) and (1.3,2.2) respectively. They yield the tightest SQSL for the considered system over all possible values of $\theta$ and $\phi$. In FIG.(b), $\frac{T}{T_{SQSL}}$ is plotted against time for different initial states, with $M=2$. The continuous lines represent the stronger bounds, given by the optimum pair of $\theta$ and $\phi$.The dashed lines correspond to the MT bounds of the same system.  The line with bigger dots represents the stronger bound corresponding to $|\Psi^\perp(t) \rangle=\frac{A-\langle A \rangle}{\Delta A}|\Psi(t)\rangle$, for the initial state $\alpha=\frac{1}{\sqrt{2}}$. The horizontal dotted line is for $\frac{T}{T_{SQSL}}= \frac{T}{T^{MT}_{QSL}}=1$.}
    \label{fig:two graphs1}
\end{figure*}
 
    Using all these expressions in Eq (\ref{35}) and (\ref{10}), $T_{SQSL}$ can be expressed in terms of $\theta$ and $\phi$. In addition to that $T/T_{SQSL}$ also depends on the evolution time, $T$. In order to avoid any explicit time dependence during the optimization, we consider the time average quantity,  $\overline{\frac{T}{T_{SQSL}}}$ instead of $\frac{T}{T_{SQSL}}$ itself.
     As shown in the Fig.3(a), the minimum value of $\overline{\frac{T}{T_{SQSL}}}$ for the initial state $|\psi(0)\rangle
    _i= \frac{1}{\sqrt{3}}|0\rangle + \sqrt{\frac{2}{3}}|1\rangle$ occurs at two pairs of $\theta$ and $\phi$: $(0.9, 4.4)$ and $(2.2, 1.3)$. The orthogonal states corresponding to these optimal pairs give the optimized bound for the above state. Following the same procedure we obtain the optimized bounds for the system with different initial states. This is depicted in the Fig.3(b) which indicates that the obtained bounds are much tighter than the corresponding MT bounds. The bound can be closely achieved for different initial states. A comparison is made regarding the performance of the two choices (\ref{21}) and (\ref{34}) for the initial state, $|\psi(0)\rangle_i=\frac{1}{\sqrt{2}}|0\rangle+\frac{1}{\sqrt{2}}|1\rangle$ in the same figure. It can be clearly seen that, the later choice of orthogonal state (\ref{34}) significantly outperforms the previous choice (\ref{21}) at least for this case. Although the above choice (\ref{34}) is valid only for two dimensional subsystems, one can find similar choice of orthogonal state for higher dimensional subsystems by considering the generalised set of Pauli matrices in the corresponding higher dimensional sub-spaces.
    
    \subsection{SQSL for Entangled State}
    In the previous section, we analysed our bound for those systems that start with a product state and remain in the product state throughout the non-interacting evolution. Here, we explore whether the bound holds tight even for the entangled systems? The evolution considered here is unitary and interaction free. Without any discrepancy we can consider a system with the same driving Hamiltonian given in Eq (\ref{14}) with an entangled initial state, say \cite{Giovannetti_2004}
    \begin{eqnarray}
    \label{40}
    |\Psi (0)\rangle=\frac{1}{\sqrt{N}}\sum_{n=0}^{N-1}|n\rangle_1 |n\rangle_2...|n\rangle_M .   
    \end{eqnarray}
    The average energy of the system is same as Eq (\ref{15}) whereas the uncertainty in energy becomes $\Delta H=\frac{\hbar\omega M \sqrt{N^2-1}}{2\sqrt{3}}$. Using the relation (\ref{2}), the MT bound for the above system can be easily obtained as
    \begin{eqnarray}
    \label{43}
    T^{MT}_{QSL}=2\sqrt{3}\frac{cos^{-1}(\left|\langle \Psi(0)|\Psi(T) \rangle \right|)}{M\sqrt{N^2-1}},
    \end{eqnarray}
     where 
    \begin{eqnarray}
    |\langle \Psi(0)|\Psi(T) \rangle|^2= \left |\frac{1}{N}\frac{sin(\frac{M \omega T N}{2})}{sin(\frac{M \omega T}{2})}\right |^2
    \end{eqnarray}
    is the transition probability between the initial and the final state. In order to evaluate the stronger bound, we use the orthogonal state as given by Eq (\ref{21}). This yields the following expression of $R(t)$  
    \begin{eqnarray}
    R(t)=\frac{1}{2}\left|1-i\frac{\langle AH\rangle-\langle A\rangle\langle H\rangle}{\Delta H\Delta A}\right|^2.
    \end{eqnarray}
     Here
    \begin{eqnarray}
     \langle AH \rangle &=\langle \Psi(t)|\Psi(0)\rangle \langle \Psi(0)|H|\Psi(t)\rangle, \nonumber
     \end{eqnarray}
     \begin{eqnarray}
     \langle \Psi(0)|H|\Psi(t)\rangle =\frac{M e^{-iM\omega t}}{N\widetilde{Z}_1}\left[\frac{1-\widetilde{Z}_2}{\widetilde{Z}_1}-(N-1)\widetilde{Z}_2\right],\nonumber
  \end{eqnarray}
   \begin{eqnarray}
    \widetilde{Z}_1 =1-e^{-iM\omega t}, \widetilde{Z}_2=e^{-iM(N-1)\omega t}. \nonumber
  \end{eqnarray}
   Using the above expressions in Eq (\ref{10}), the SQSL for the above system can be calculated. In Fig.4, we plot the stronger bound and the MT bound for the system.
   
    \par 
    As shown in the Fig.4, the new bound remains consistently tight with the evolution time for any value of $N$. In contrast to that, MT bound slowly loosens up as evolution time increases. The increasing $\delta$ with increasing time indicates the new bound significantly outperforms the MT bound at greater evolution times. In the Fig.5, the bound is observed for different values of $M$. The new bound remains tighter than the MT bound for any value of $M$. With the increasing $N$, the new bound significantly outperforms the MT bound whereas the performance nearly same for all the considered values of $M$. Thus, the stronger bound suitably works for the entangled system(\ref{40}) with the considered choice of orthogonal state.
        \par
    Proceeding in a similar way as in the Section I, we intend to obtain an expression like (\ref{35}) for the entangled systems. Consider an initial state of the form
    \begin{eqnarray}
    \label{47}
    |\Psi(0)\rangle=\alpha |0\rangle_1 |0\rangle_2 ... |0\rangle_M + \beta |1\rangle_1 |1\rangle_2 ... |1\rangle_M   
    \end{eqnarray}
    evolving under the same Hamiltonian as mentioned in (\ref{14}), where $N=2$. The system has the average energy  $\langle H \rangle = M |\beta|^2 \omega \hbar$ and the uncertainty in energy $\Delta H = M \omega \hbar |\alpha| |\beta|$. Using (\ref{2}) and choosing $\omega, \hbar=1$ the MT bound for the above system can be written as
    \begin{eqnarray}
    \label{50}
    T_{QSL}^{MT}=\frac{\cos^{-1}{|\langle\Psi(0)|\Psi(T)\rangle|}}{M |\alpha||\beta|},
    \end{eqnarray}
    where the transition probability at any time,
    \begin{eqnarray}
    \label{49}
    \begin{aligned}
    |\langle\Psi(0)|\Psi(t)\rangle|^2=|\alpha|^4+|\beta|^4+2 |\alpha|^2 |\beta|^2 \cos({M\omega t}).
     \end{aligned}
    \end{eqnarray}
    As for the orthogonal state we consider the same expression as (\ref{34}) which yields an expression of $R(t)$ same as (\ref{35}). Here
    \begin{eqnarray}
    \label{51}
    \langle O \rangle &=\widetilde{ A}_1\cos^M{\theta}+\widetilde{ A}_2\sin^M{\theta},
    \end{eqnarray}
   \begin{eqnarray}
    \label{53}
    \begin{aligned}
    \langle O A \rangle & = \widetilde{B}_1\cos^M\theta+\widetilde{B}_2\sin^M\theta,
     \end{aligned}
   \end{eqnarray}
    \begin{eqnarray}
    \label{54}
    \langle O H \rangle=\widetilde{C}_1\cos^M{\theta}+\widetilde{C}_2\sin^M{\theta},
    \end{eqnarray}
    and 
    \begin{eqnarray}
    \Delta O =\sqrt{1-\langle O \rangle^{2}}.\nonumber
     \end{eqnarray}
        \begin{figure}[htp]
    \centering    
    \includegraphics[width=8cm]{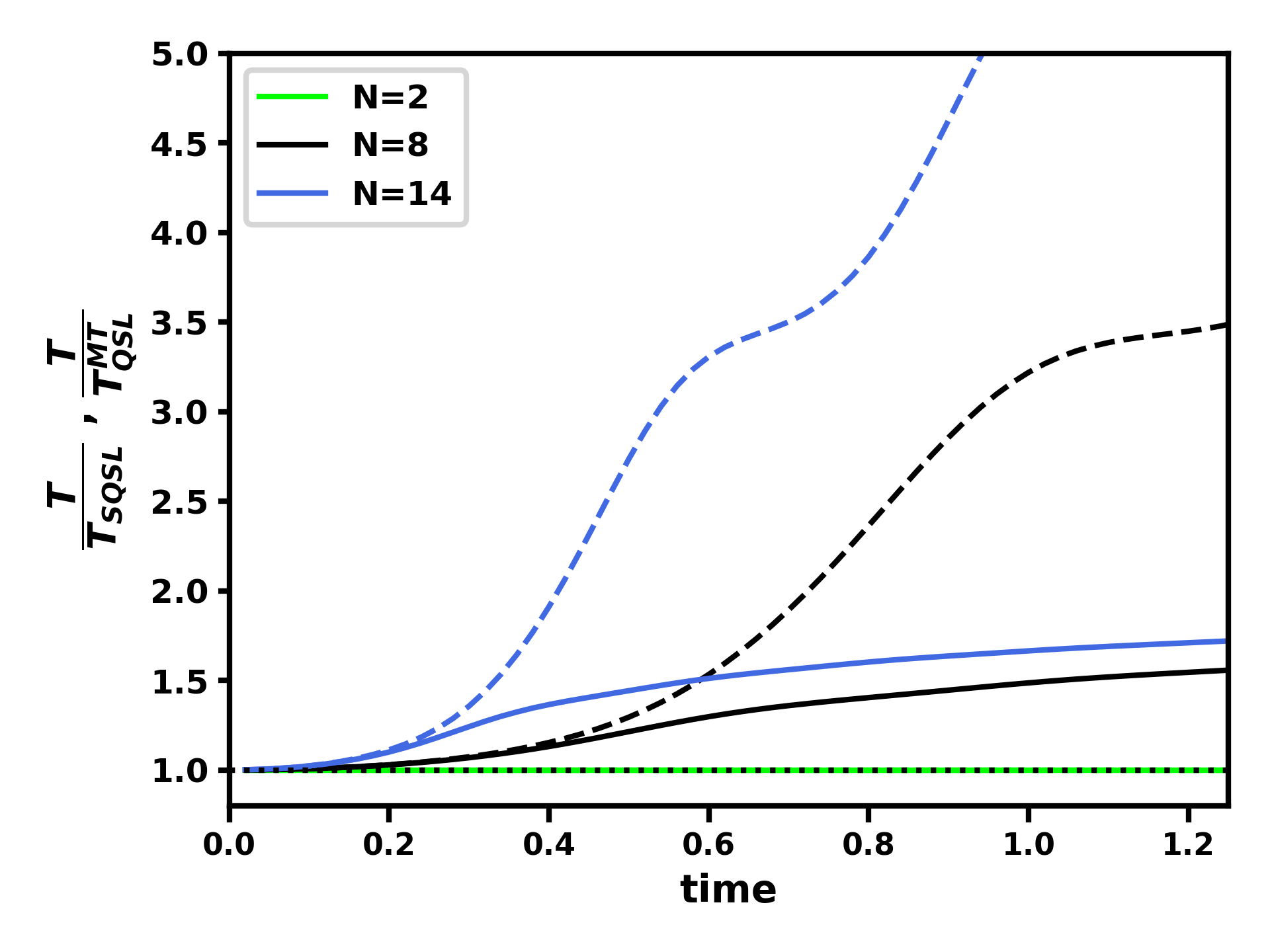}
    \caption{\(T/T_{SQSL}\) and \(T/T^{MT}_{QSL}\)are plotted against time for different values of $N$ of the entangled system, by fixing the value of $M$ as \(M=2\). The MT bounds are represented by the dashed lines and the strong bounds are represented by the continuous lines. The dotted line represents $T/T_{SQSL}=T/T^{MT}_{QSL}=1$.}
    \label{fig:4}
    \end{figure}
    \begin{figure}[htp]
    \centering    \includegraphics[width=8cm]{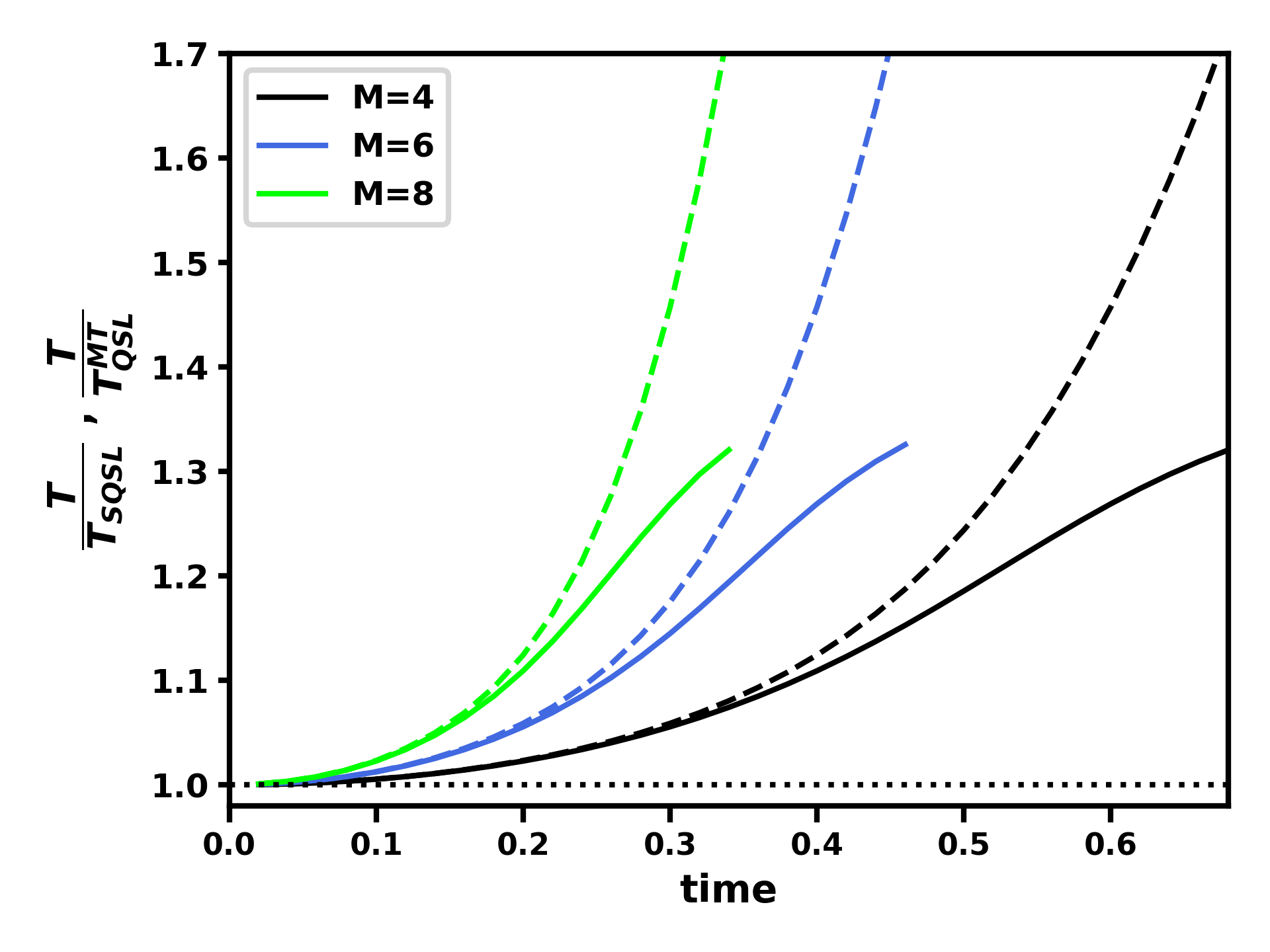}
    \caption{\(T/T_{SQSL}\) and \(T/T^{MT}_{QSL}\)are plotted against time for different values of $M$ by fixing the value of $N$ as \(N=2\). The MT bounds are represented by the dashed lines and the new bounds are represented by the continuous lines. The straight dotted line represents $T/T_{SQSL}=T/T^{MT}_{QSL}=1$.}
    \label{fig:5}
    \end{figure}
    The terms $\widetilde{A}_1$, $\widetilde{A}_2$, $\widetilde{B}_1$, $\widetilde{B}_2$, $\widetilde{C}_1$, and $\widetilde{C}_2$ can be expressed as
    \begin{align*}
    \widetilde{A}_1 &=|\alpha|^2+(-1)^M 
    |\beta|^2,\quad \widetilde{A}_2 =2 Re(\alpha \beta^*e^{iM(\omega t+\phi)}),\\ \widetilde{B}_1 &=|\alpha|^4+(-1)^M |\beta|^4+|\alpha|^2|\beta|^2(e^{-iM\omega t}+(-1)^M e^{iM\omega t}),\\
    \widetilde{B}_2 &=(\alpha ^*\beta e^{-iM\phi}+\alpha \beta^*e^{iM(\omega t+\phi)})(|\alpha|^2+|\beta|^2 e^{-iM \omega t}),\\ \widetilde{C}_1 &=M (-1)^M|\beta|^2 ,\quad \widetilde{C}_2 =M \beta \alpha^* e^{-i M(\phi +  \omega t)}\nonumber.
    \end{align*}
    Using the above expressions along with (\ref{10}), the expression for $T_{SQSL}$ can be obtained. Applying the optimization procedure as described for the separable state, we obtain the optimized bounds for different initial states (\ref{47}) of the entangled system.

\begin{figure*}
 \centering    
    \includegraphics[width=16cm]{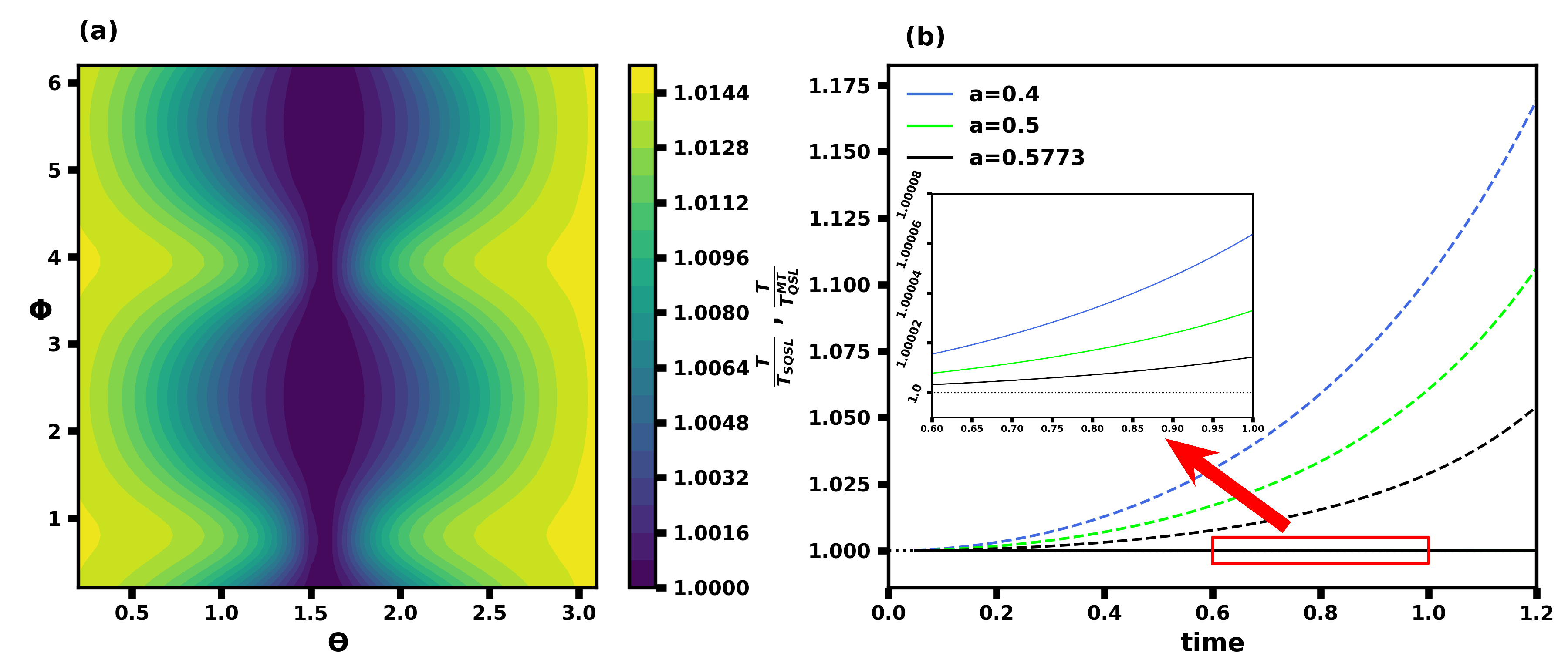}
     \caption{In Fig.(a), the initial bipartite state with $\alpha=\frac{1}{\sqrt{3}}$ has been considered. The estimated values of $\overline{\frac{T}{T_{SQSL}}}$ for different values of $\theta$ and $\phi$ have been depicted in this figure. The dark blue region represents the optimal values of $\theta, \phi$ which may yield the highly tight SQSL bounds. The yellow region on the other hand represents the less tight SQSL bounds. In Fig.(b), the ratio  $\frac{T}{T_{SQSL}}$ is plotted against time for different initial  states, with $M=2$. The dashed lines represent the MT bounds whereas the continuous lines represent the stronger bounds, given by the optimum pair of $\theta$ and $\phi$ corresponding to each state. The horizontal dotted line represents $\frac{T}{T_{SQSL}}= \frac{T}{T^{MT}_{QSL}}=1$.}
    \label{fig:two graphs2}
\end{figure*}


In Fig.8(a) we see that the minimum value of $\overline{\frac{T}{T_{SQSL}}}$ occurs almost at all values of
 $\phi$ for certain values of $\theta$. Therefore, the expression (\ref{34}) suggests many orthogonal states that yield optimum bound for the system. The optimum bounds are plotted for different initial states in the Fig.8(b). For any of the considered states, the average ratio $\overline{\frac{T}{T_{SQSL}}}$ becomes very close to 1. For instance, for $\alpha=\frac{1}{\sqrt{3}}$ and $\beta=\sqrt{\frac{2}{3}}$, $\overline{\frac{T}{T_{SQSL}}}\approx 1.00001$. Thus, the curves corresponding to the stronger bound closely overlap with the line representing $T/T_{SQSL}=1$. The MT bounds on the other hand, maintain appreciable distance from the same line. This depicts the tightness of the new bound. Undoubtedly, this form of orthogonal states (\ref{34}) performs extremely well for the considered entangled system.
  \subsection{SQSL for Interacting systems}
  Till now we have studied the bound (\ref{10}) in the context of non-interacting evolution, where the states evolve by local unitary operations. In this case every subsystems evolve independently, without affecting each-other's evolution. The study in the previous sections show that the bound holds tight for such evolution. What about an interaction assisted evolution? In general when systems interact, their evolution may get affected by each other. During the evolution, correlations like entanglement may get established or degraded due to interaction. Thus, interaction may either fasten up or slower the system's evolution. Does the bound perform effectively in such scenario? To illustrate this we consider a simple model as described in Ref. \cite{Giovannetti_2004}. Consider a chain of $M$  spins that evolve under the Hamiltonian 
   \begin{eqnarray}
   \label{55}
    H=\sum_{i=1}^{M} H_i+H_{int},
   \end{eqnarray}
    where
   \begin{eqnarray}
   H_i&=\hbar \omega_0 (1-\sigma_x ^i).\nonumber \end{eqnarray}
   The local Hamiltonian, $H_i$ evolves the individual systems independently, whereas $ H_{int} $ acts on subsystems together. According to the considered model, let us assume the interaction takes place in each of the $Q$ number of blocks present in the spin chain. Each block consists of $K$ number of spins, $K<M$. The $K$ spins in the $j^{th}$ block interact through the Hamiltonian $H_j$\quad : 
   \begin{align*}
    H_j=\hbar \omega (1-S_j),\\
    where\quad  S_j=\sigma_x^{i_{1j}}\otimes \sigma_x^{i_{2j}}\otimes....\otimes \sigma_x^{i_{kj}},\\
    j= 1, 2,..., Q.
    \end{align*}
    Therefore, the total interaction Hamiltonian becomes
    \begin{align*}
    H_{int}&=\hbar \omega \sum_{j=1}^{Q} (1-S_j).
    \end{align*}
    The system initially is in a product state 
   $|\Psi(0)\rangle=|0\rangle_1 |0\rangle_2 .... |0\rangle_M$, where $|0\rangle$ is the eigen state of the Pauli-z matrix, $\sigma_z$.  If we make it evolve under the Hamiltonian (\ref{55}), then the evolved state becomes \cite{Giovannetti_2004} 
   \begin{eqnarray}
   \begin{aligned}
   |\Psi (t)\rangle&=C\bigotimes_{i=1}^{M}(\cos{\omega_0 t}+i\sigma_x^{i}\sin{\omega_0 t})
   \prod_{j=1}^{Q}(\cos{\omega t}\\&+iS_j\sin{\omega t})|\Psi (0)\rangle,
    \end{aligned}
   \end{eqnarray}
   where $C=e^{-i(\omega+\omega_0)t}$. A further simplified picture can be obtained by imposing the condition that each and every spin gets involved in only two of the $Q$ interaction blocks. This establishes a relation between the number of blocks, $Q$ and the order of interaction, $K$ as $Q=\frac{2M}{K}$.  Since the interaction under consideration, $S_j$ reverts the spins in the $jth$ block, the consecutive action of all the $S_j$ s at a time leaves no change in the spin states, i.e., 
   $S_1 S_2 ....S_Q=1$.

  Using the above condition the transition probability become
  \begin{eqnarray}
  \label{56}
    |\langle \Psi(t)|\Psi(0)\rangle|^2 &= |(\cos{\omega t})^Q+(i \sin{\omega t})^Q|^2.
    \end{eqnarray}
    Previous study \cite{Giovannetti_2003} shows that the strong interaction, $\omega>> \omega_0$ helps the model to achieve the unified bound \cite{Levitin2009} by building up sufficient correlations among the sub-systems. This encourages us to study our bound in the regime of strong interaction. In this regime, the average energy for the above system is given by $\langle H \rangle = M \omega_0 + Q \omega \approx Q \omega$
     and the fluctuation in energy $
     \Delta H = \sqrt{M \omega_0 ^2 + Q \omega^2} \approx \sqrt{Q} \omega$.
   Thus, the MT bound in this regime can be written as
    \begin{equation}
    \label{59}
    T^{MT}_{QSL}=\frac{\cos^{-1}{|(\cos{\omega t})^Q+(i \sin{\omega t})^Q|}}{\sqrt{Q} \omega}. 
    \end{equation}
    In order to obtain the stronger bound (\ref{10}) for the system, the orthogonal state $|\Psi^{\perp}(t)\rangle$ has to be specified. Previously, we have discussed certain forms of $|\Psi^{\perp}(t)\rangle$ which yield tighter bound in case of the non-interacting evolution. It is therefore, interesting to check how well they capture an interacting evolution. We have calculated the SQSL for different choices of $|\Psi^{\perp}(t)\rangle$. First we can check for 
    $|\Psi^{\perp}(t)\rangle=\frac{A-\langle A \rangle }{\Delta A}|\Psi(t)\rangle$,
    where
    $A=|\Psi(0)\rangle\langle\Psi(0)|$. For this choice of orthogonal state, $R(t)$ is described by the Eq. (\ref{22}), where 
    \begin{align*}
    \langle AH\rangle &=\kappa [\cos{\omega t}(i\sin{\omega t})^{Q-1}+i \sin{\omega t}(\cos{\omega t})^{Q-1}],\\
   \kappa &=-\omega Q[(\cos{\omega t})^Q+(i \sin{\omega t})^Q].
    \end{align*}
       \begin{figure}[t]
    \centering
    \includegraphics[width=8cm]{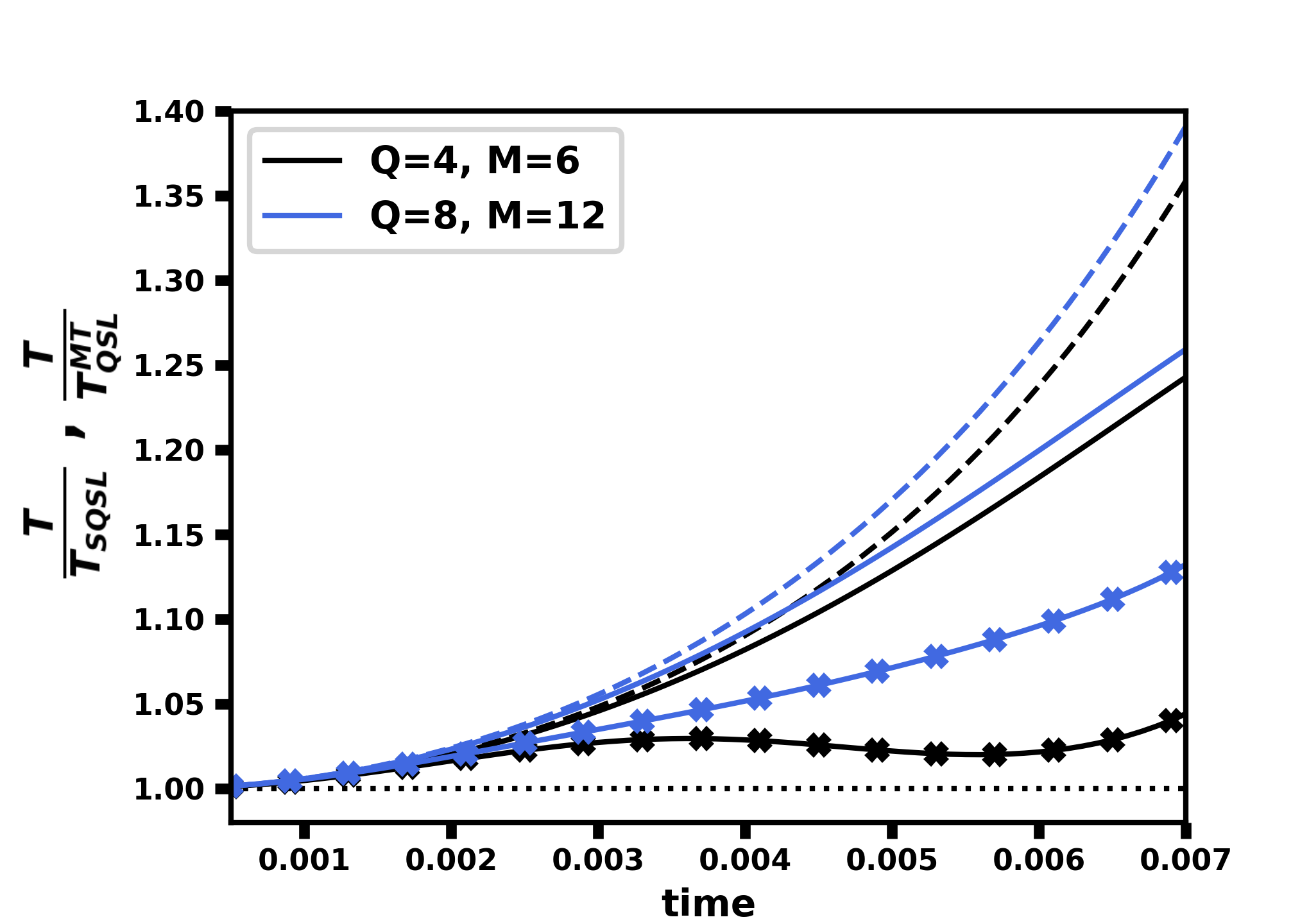}
    \caption{\(T/T_{SQSL}\)
     and \(T/T^{MT}_{QSL}\) vs time are plotted for two different systems represented by different pairs of $Q$ and $M$. The MT bounds are represented by the dashed lines and the stronger bounds corresponding to $|\Psi^{\perp}(t)\rangle=\frac{A-\langle A \rangle }{\Delta A}|\Psi(t)\rangle$ are represented by the continuous lines. Again the crossed lines represent the SQSLs with $|\Psi^{\perp}(t)\rangle=\frac{O-\langle O \rangle }{\Delta O}|\Psi(t)\rangle$.  The straight dotted line represents $T/T_{SQSL}=T/T^{MT}_{QSL}=1$.}
    \label{fig:7}    
    \end{figure}
    Using the above expressions along with (\ref{56}) we derive $T_{SQSL}$ for the system. Second we can check 
     $|\Psi^{\perp}(t)\rangle=\frac{O-\langle O \rangle }{\Delta O}|\Psi(t)\rangle$,
     where $O$ is given by Eq. (\ref{34}). For this choice the expression of $R(t)$ is given by Eq. (\ref{35}).We estimate $T_{SQSL}$ for this choice of orthogonal state using analytical as well as computational tools.\par
  
    The estimated bounds are depicted in Fig.7. Here, two systems with different number of sub-systems, $M$ and different orders of interaction, $K$ are considered. For each case, the stronger bound (SQSL) and the MT bound are observed. The SQSLs of the respective systems yield tighter bounds compared to the corresponding MT bounds.
    Again, for each system, the choice   $|\Psi^{\perp}(t)\rangle=\frac{O-\langle O \rangle }{\Delta O}|\Psi(t)\rangle$ among all other choices, yields the tightest bound in the observed time interval. Undoubtedly, our bound well captures the evolution time of the systems undergoing interacting dynamics.  

    \section{Conclusions}
    The Heisenberg-Robertson uncertainty relation is one of the fundamental result in early days of development of quantum theory. Even though, this cannot be used directly to derive the time-energy uncertainty relation, nevertheless, it provided a firm ground to prove
    the quantum speed limit such as the MT bound. In this paper, we have proved stronger quantum speed limit (SQSL) for arbitrary unitary evolution of quantum systems in the pure states. This is possible with the help of the stronger uncertainty relations in quantum theory. We have shown that the MT bound follows as a special case of stronger quantum speed limit. We have applied the new bound for composite systems prepared in separable state as well as in the entangled state. We find that the new bound provides accurate time that the system takes to evolve from an initial state to a final state compared to MT bound, where the later bound is very loose.
    We have also applied the new bound for interacting system and find that with the suitable choice of the orthogonal state, the SQSL can surpass the MT bound. We believe that our findings have important application in a wide variety of physical systems ranging from optimal control to fast implementation of logic gates in quantum computers and alike. Moreover, the SQSL can be experimentally tested with few qubit systems which are already available with near term devices.
    
    \section{ACKNOWLEDGMENTS}
D. Thakuria acknowledges the financial support from Quantum Enabled Science and Technology (QuEST) project from the Department of Science and Technology, India. A. K. Pati acknowledges the support from the J C Bose grant from the Department of Science and Technology, India. D.Thakuria also acknowledges B. Mohan and S.Talukdar for useful discussions. 
    
    \bibliography{qsl.bib}
    

 \end{document}